\documentclass[10pt, conference, letterpaper]{IEEEtran}
\usepackage[letterpaper, left = 0.68in, right =0.68in, top = 0.7in, bottom = 1in]{geometry}
\def\arxiv{1}
\def\arxivdisclaimer{1} %
\def\review{0} %

\usepackage[noadjust]{cite}
\PassOptionsToPackage{svgnames}{xcolor} %
\if\review1
\usepackage{todonotes}
\else
\usepackage[disable]{todonotes}
\fi
\usepackage[acronym]{glossaries}
\usepackage{mathtools} %
\usepackage{amsthm}
\usepackage{amsmath, amssymb}
\usepackage{balance}

\usepackage{svg} %
\usepackage[svgnames]{xcolor} %
\usepackage{graphicx} %
\usepackage{pgfplots} %
\usepgfplotslibrary{groupplots} %
\usetikzlibrary{spy} %
\usetikzlibrary{arrows.meta} %
\usepackage{xfrac} %
\usepackage{siunitx} %
\pgfplotsset{compat=newest} %
\pgfplotsset{
    /pgfplots/layers/Bowpark/.define layer set={
        axis background,axis grid,main,axis ticks,axis lines,axis tick labels,
        axis descriptions,axis foreground
    }{/pgfplots/layers/standard},
    colormap={jet}{
        rgb255(0cm)=(0,0,128);
        rgb255(1cm)=(0,0,255);
        rgb255(3cm)=(0,255,255);
        rgb255(5cm)=(255,255,0);
        rgb255(7cm)=(255,0,0);
        rgb255(8cm)=(128,0,0)
    }
} %

\if\arxiv0
\else
\usepackage{hyperref} %

\glsdisablehyper
\fi
\usepackage{cleveref}
\usepackage{subcaption}
\usepackage{makecell} %
\usepackage{algorithm} %
\usepackage{algpseudocode} %
\usepackage{lipsum} %
\usepackage{textcase}
\usepackage[tablename=TABLE,font=small]{caption}
\DeclareCaptionTextFormat{up}{\MakeTextUppercase{#1}}
\newcommand{\hn}{\mkern-1.5mu}

\newcommand\blfootnote[1]{%
  \begin{NoHyper}%
  \begingroup
  \renewcommand\thefootnote{}\footnote{#1}%
  \addtocounter{footnote}{-1}%
  \endgroup
  \end{NoHyper}%
}

\newtheorem*{remark*}{Remark}

\begin{document}
\title{BASIIS: Bistatic Angular Sampling and Interpolation for ISAC Setups}

\author{

\IEEEauthorblockN{
        Alexander~Felix\IEEEauthorrefmark{2}\IEEEauthorrefmark{1},
        Marcus~Henninger\IEEEauthorrefmark{1}, 
        Lucas~Giroto\IEEEauthorrefmark{1}, 
        Maximilian~Bauhofer\IEEEauthorrefmark{2}\IEEEauthorrefmark{1},\\
        Stephan~ten~Brink\IEEEauthorrefmark{2},
        and Silvio Mandelli\IEEEauthorrefmark{1} %
        }
        
\IEEEauthorblockA{
\IEEEauthorrefmark{1}Nokia Bell Labs Stuttgart, 70469 Stuttgart, Germany \\
\IEEEauthorrefmark{2}Institute of Telecommunications, University of Stuttgart, 70569 Stuttgart, Germany \\ 
\if\arxivdisclaimer1
E-mail: \href{mailto:felix@inue.uni-stuttgart.de}{felix@inue.uni-stuttgart.de}} \vspace{-5mm} 
\else
E-mail: felix@inue.uni-stuttgart.de
\fi
\thanks{TBD: Further notes.}}

\maketitle

\newacronym{1D}{1D}{one-dimensional}
\newacronym{2D}{2D}{two-dimensional}
\newacronym{3D}{3D}{three-dimensional}
\newacronym{3GPP}{3GPP}{3rd Generation Partnership Project}
\newacronym{4D}{4D}{four-dimensional}
\newacronym{5G}{5G}{fifth generation}
\newacronym{6G}{6G}{sixth generation}
\newacronym{agcd}{AGCD}{approximate greatest common divisor}
\newacronym{awgn}{AWGN}{additive white Gaussian noise}
\newacronym{cfar}{CFAR}{constant false alarm rate}
\newacronym{dft}{DFT}{discrete Fourier transform}
\newacronym{gcd}{GCD}{greatest common divisor}
\newacronym{if}{IF}{interpolation factor}
\newacronym{isac}{ISAC}{Integrated Sensing and Communications}
\newacronym[plural={MRAs}]{mra}{MRA}{minimum redundancy array}
\newacronym{lmf}{LMF}{Location Management Function}
\newacronym{mf}{MF}{management function}
\newacronym{naf}{NAF}{normalized angular frequency}
\newacronym{NRPPa}{NRPPa}{NR Positioning Protocol A}
\newacronym{poc}{PoC}{proof of concept}
\newacronym[plural={PSFs}]{psf}{PSF}{point spread function}
\newacronym[plural={RUs}]{ru}{RU}{radio unit}
\newacronym{rx}{RX}{receiver}
\newacronym{rmse}{RMSE}{root mean squared error}
\newacronym{sar}{SAR}{synthetic-aperture radar}
\newacronym{sara}{SARA}{Sampling and Reconstructing Angular Domains with Uniform Arrays}
\newacronym{semf}{SeMF}{Sensing Management Function}
\newacronym{tx}{TX}{transmitter}
\newacronym[plural={TRPs}]{trp}{TRP}{transmission reception point}
\newacronym[plural={ULAs}]{ula}{ULA}{uniform linear array}
\newacronym[plural={URAs}]{ura}{URA}{uniform rectangular array}

\begin{abstract}
\acrfull{isac} is a defining feature of 6G, extending cellular networks with radar-like sensing at limited additional overhead. In bistatic deployments, sensing requires coordinating the \acrfull{tx} and \acrfull{rx} arrays 
to scan the Cartesian product of angle of departure and arrival, resulting in a four-dimensional sampling problem in the angular domain.

This work establishes a complete angular sampling framework for bistatic \acrshort{isac}, extending the \acrshort{dft}-based optimal-sampling methodology to the full azimuth and elevation domains of both arrays. We show that the bistatic geometry couples the \acrshort{tx} and \acrshort{rx} elevation angles, and represent this coupling through the ortho-baseline coarray, a virtual array that captures the joint elevation aperture of the array pair.
From the coarray we derive a minimal sampling and interpolation scheme, near-lossless and realizable with any beamforming architecture. Monte Carlo simulations confirm the proposed minimal acquisition essentially equalizes the detection accuracy of dense oversampled imaging while acquiring $3$ to $5$ times fewer \acrshort{tx}--\acrshort{rx} direction pairs. This allows having bistatic operations with drastically reduced overhead on the radio resource usage of \acrshort{isac} systems.
\end{abstract}

\if\arxivdisclaimer1
\blfootnote{
This work has been submitted to the IEEE for possible publication. Copyright may be transferred without notice, after which this version may no longer be accessible.
}
\vspace{-5mm}
\else
\vspace{1mm}
\vspace{-2mm}
\fi

\IEEEpeerreviewmaketitle

\glsresetall
\section{Introduction}
\gls{isac} is a defining feature of the upcoming \gls{6G} of cellular communications, targeting \gls{3GPP} Release 20~\cite{3GPP_TR_22870}.
\gls{isac} aims to extend the existing communications infrastructure with radar-like capabilities, giving the network awareness of its passive surroundings with limited additional overhead~\cite{ghosh2025unified}.

A compressed scene representation, such as an estimate of the direction of a single target, is sufficient for \gls{isac} use cases (e.g., drone detection). By contrast, area monitoring requires the full scene response in azimuth and elevation to perform an imaging task~\cite{manzoni2025wavefield}, one example being automated guided vehicles moving in a factory hall~\cite{3GPP_TR_22837}. Brown-field deployment favors bistatic operation: monostatic sensing needs full-duplex operation, limited by self-interference~\cite{girotoBistaticISAC2026}, or a second co-located array, whereas bistatic pairs reuse existing sites with overlapping coverage. We address the imaging task in the angular domain, where the \textit{image} is the scene response over the independent angular components.

Obtaining the image requires solving the angular sampling task: identifying the set of \gls{tx} and \gls{rx} direction pairs to be acquired by beamforming, such that every non-sampled direction of interest can be recovered with the minimal number of pairs. Its scope depends on the topology: one angular component for monostatic operation in a single angular domain, but four when both arrays scan azimuth and elevation in a bistatic setup. Of these four, the image retains three, since both beams must meet in a common scan point, which ties the \gls{tx}--\gls{rx} elevation angles. Exhaustively acquiring the direction pairs is therefore impractical: the acquisition time grows as the product of the per-component direction counts, and analog and hybrid beamforming architectures further restrict achievable sampling~\cite{rajamaki2019analog, rajamaki2020hybrid}. Minimizing this acquisition is essential, all the more for systems with limited angular capabilities, distributed \gls{isac} with constrained fronthaul, and communications-sensing coexistence on shared angular resources.

The optimal angular sampling task is addressed in prior art for monostatic setups~\cite{mandelli2022sampling, treesOptimumArrayProcessing2002} and for azimuth-only operation of bistatic setups~\cite{felixOptimalAzimuthSampling2025a}. Bistatic and multistatic \gls{isac} platforms, by contrast, still acquire the angular domain by sweeping dense beam codebooks~\cite{paidimarriEyeBeamMmWave5GCompliant2024} or with fixed beams~\cite{beusterEnhancingSituationalAwareness2025}, and the MIMO-OFDM system concept of~\cite{girotoBistaticMimoSync2025} addresses synchronization and processing rather than the beam pairs to acquire. To the best of the authors' knowledge, no minimal sampling solution covers the bistatic elevation domain, in which the requirement that both beams meet in a common scan point couples the \gls{tx} and \gls{rx} angles and removes the per-array separability exploited in the azimuth-only case. Extending minimal sampling to the full \gls{4D} bistatic angular domain is the goal of this work.

We propose \emph{BASIIS} (Bistatic Angular Sampling and Interpolation for \gls{isac} Setups) to address the above challenges by:
\begin{itemize}
    \item deriving how the \gls{4D} bistatic scan space collapses onto three independent angular components, characterizing the joint elevation aperture of the array pair as an ortho-baseline coarray for each azimuth pair,
    \item solving, on this basis, the angular sampling task of generic bistatic setups with a minimal Nyquist--Shannon sampling set and a Dirichlet kernel interpolation, extending~\cite{mandelli2022sampling, felixOptimalAzimuthSampling2025a} for near-lossless reconstruction,
    \item evaluating BASIIS in Monte Carlo simulations against dense oversampled references, with matched detection accuracy at $3.1$ to $5.3\times$ fewer acquired direction pairs.
\end{itemize}

\section{System Model} \label{sc:model}

\begin{figure}[t]
  \centering
  {
    \def\svgwidth{1.0\columnwidth}
    \def\h{\bf}
    \def\d{}
    \def\f{\small\tt}
    \def\l{\footnotesize}
    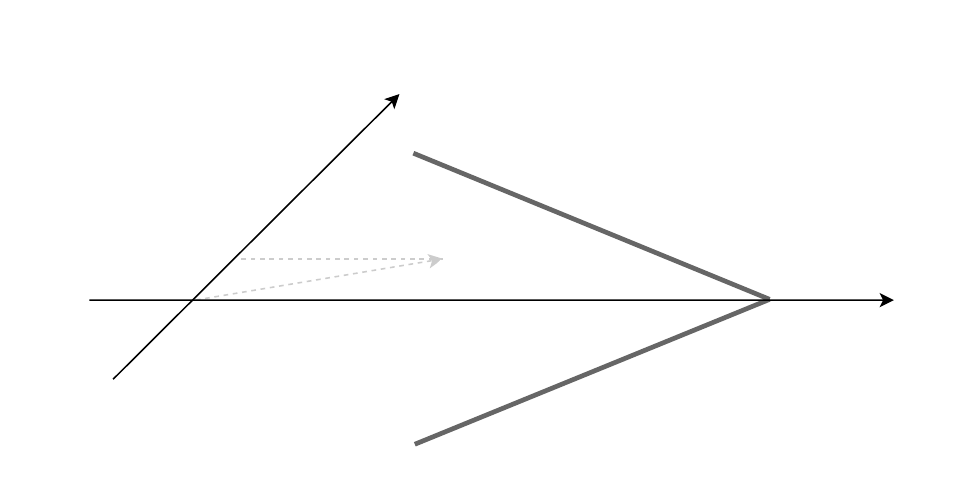
    \caption{Bistatic scan point on the circumference (red) of radius $r_c$, where the azimuth cones intersect for a fixed pair $(\vartheta^{(\mathrm{t})},\vartheta^{(\mathrm{r})})$. The circumference is centered at $x^\star$ on the baseline with length $b$, and the scan point couples the elevation pair $(\varphi^{(\mathrm{t})},\varphi^{(\mathrm{r})})$.}
    \label{fig:bistatic_ang_pairs}
  }
\end{figure}

We consider a static bistatic setup~\cite{willisBistaticRadar2005} consisting of spatially separated \gls{tx} and \gls{rx} arrays, as in Fig.~\ref{fig:bistatic_ang_pairs}. The spatial separation of length $b$ is the \textit{bistatic baseline}. Without loss of generality, we define our coordinate system such that the bistatic baseline is aligned with the $x$-axis. The direction perpendicular to the baseline within the reference array plane is referred to as the \textit{ortho-baseline}, here the $z$-axis. The \gls{tx} is located at $\left(0,0,0\right)$ and the \gls{rx} at $\left(b,0,0\right)$. Throughout, a superscript $(\hn\cdot\hn)$ marks a quantity specific to each array, replaced by $(\mathrm{t})$ for the \gls{tx} or $(\mathrm{r})$ for the \gls{rx}. Quantities written without it are common to both arrays. Array elements lie in the $x$-$z$ plane, with positions $\mathbf{p}_i^{(\hn\cdot\hn)} = [x_i^{(\hn\cdot\hn)}, 0, z_i^{(\hn\cdot\hn)}]^\intercal$. Array rotations, not considered here, enter only through the element positions projected onto the plane transverse to the baseline.
The \gls{tx} and \gls{rx} are \glspl{ura} with $N_x^{(\hn\cdot\hn)}$ elements along the baseline and $N_z^{(\hn\cdot\hn)}$ along the ortho-baseline, so $N^{(\hn\cdot\hn)}\!=\!N_x^{(\hn\cdot\hn)}\!N_z^{(\hn\cdot\hn)}$.

\begin{figure}[t]
  \centering
  {
    \def\svgwidth{0.90\columnwidth}
    \def\h{\bf}
    \def\d{}
    \def\f{\small\tt}
    \def\l{\footnotesize}
    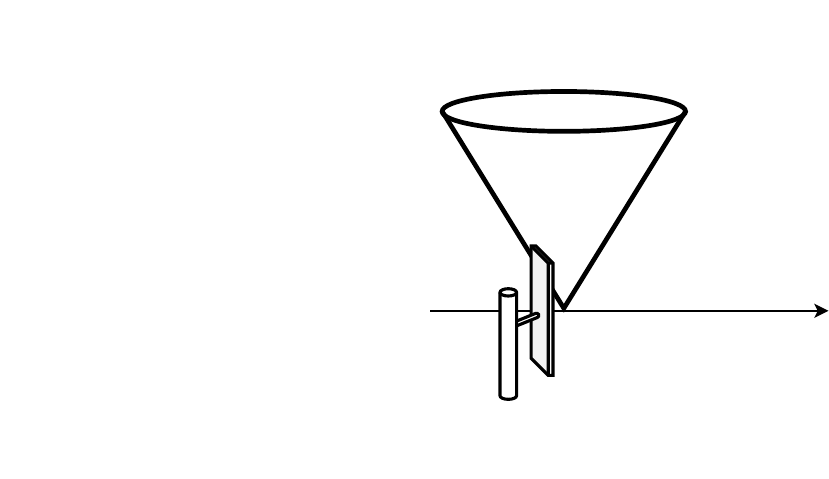
    \caption{Azimuth (left) and elevation (right) cones. A conic angle fixes a whole cone of directions, all with the same projection onto the array axis and hence the same \gls{ula}-like response.}
    \label{fig:bistatic_cones}
  }
\end{figure}

We take the \gls{rx} array, with boresight along the $y$-axis, as the reference when defining the angular sampling task.
An arbitrary beamforming direction is the unit vector
\begin{equation}
\mathbf{u}^{(\hn\cdot\hn)}= \left[\sin\vartheta^{(\hn\cdot\hn)}, \sqrt{1-\sin^2\vartheta^{(\hn\cdot\hn)}-\sin^2\varphi^{(\hn\cdot\hn)}}, \sin\varphi^{(\hn\cdot\hn)}  \right]^{\intercal} \; ,
\label{eq:DirectionVector}
\end{equation}
with the conic azimuth and elevation angles $\vartheta$ and $\varphi$ of Fig.~\ref{fig:bistatic_cones}. 
 For brevity we write $\mathbf{u}^{(\hn\cdot\hn)}$ for $\mathbf{u}^{(\hn\cdot\hn)}\left(\vartheta^{(\hn\cdot\hn)}\hn,\varphi^{(\hn\cdot\hn)}\right)$.
The key property of conic angles is that $[\mathbf{u}^{(\hn\cdot\hn)}]_x$ depends only on $\vartheta^{(\hn\cdot\hn)}$ and $[\mathbf{u}^{(\hn\cdot\hn)}]_z$ only on $\varphi^{(\hn\cdot\hn)}$. Real-valued directions require $\sin^2\vartheta^{(\hn\cdot\hn)} + \sin^2\varphi^{(\hn\cdot\hn)} \le 1$, the conic-angle visible region.

For the phase term of element $i$, with the free-space wavenumber $k_0 = 2\pi/\lambda$, we have 
\begin{equation}
    e^{-jk_0\,\mathbf{u}^{(\hn\cdot\hn)\intercal}\mathbf{p}_i^{(\hn\cdot\hn)}}
    = e^{-jk_0\!\left(x_i^{(\hn\cdot\hn)}\sin\vartheta^{(\hn\cdot\hn)} + z_i^{(\hn\cdot\hn)}\sin\varphi^{(\hn\cdot\hn)}\right)},
    \label{eq:ElementPhase}
\end{equation}
where the $y$-component contributes no phase, as elements lie in the orthogonal array plane, leaving two scalar projections.

Consider a bistatic scene described by the scatterer amplitude function
$a\!\left(\mathbf{s}^{(\mathrm{t})},\mathbf{s}^{(\mathrm{r})}\right)$,
where $\left(\mathbf{s}^{(\mathrm{t})},\mathbf{s}^{(\mathrm{r})}\right) \in \Psi$
are the \gls{tx} and \gls{rx} direction pairs of each scatterer and $\Psi$
denotes the set of all such pairs in the scene. Beamforming at the scan direction pair $\left(\hn\mathbf{u}^{(\mathrm{t})}\hn, \mathbf{u}^{(\mathrm{r})}\hn\right)$, the \gls{rx} angular response is \vspace{-4.5mm}
\begin{multline}
\hat{a}\!\left(\mathbf{u}^{(\mathrm{t})}, \mathbf{u}^{(\mathrm{r})}\right)
= \sum_{n=1}^{N^{(\mathrm{t})}}\sum_{m=1}^{N^{(\mathrm{r})}}
  w_n^{(\mathrm{t})} w_m^{(\mathrm{r})}\,
  e^{-jk_0(\mathbf{u}^{(\mathrm{t})\intercal}\mathbf{p}_n^{(\mathrm{t})} + \mathbf{u}^{(\mathrm{r})\intercal}\mathbf{p}_m^{(\mathrm{r})})} \\
\cdot\iint_\Psi a\!\left(\mathbf{s}^{(\mathrm{t})},\mathbf{s}^{(\mathrm{r})}\right)
  e^{jk_0(\mathbf{s}^{(\mathrm{t})\intercal}\mathbf{p}_n^{(\mathrm{t})} + \mathbf{s}^{(\mathrm{r})\intercal}\mathbf{p}_m^{(\mathrm{r})})}
  d\mathbf{s}^{(\mathrm{t})} d\mathbf{s}^{(\mathrm{r})},
\label{eq:BeamformedSignalBistaticRx}
\end{multline}
where $w_n^{(\mathrm{t})}$ and $w_m^{(\mathrm{r})}$ denote the beamforming coefficients of the $n$-th \gls{tx} and $m$-th \gls{rx} element, respectively~\cite{felixOptimalAzimuthSampling2025a}.

Normalizing by the per-array element spacings, $d_x^{(\hn\cdot\hn)}$ along the baseline and $d_z^{(\hn\cdot\hn)}$ along the ortho-baseline, as well as the wavelength $\lambda$, defines the \gls{naf} components~\cite{mandelli2022sampling} as \vspace{-3.5mm}
\begin{align}
    \ell^{(\hn\cdot\hn)} = \frac{d_x^{(\hn\cdot\hn)}}{\lambda}\,\left[\mathbf{u}^{(\hn\cdot\hn)}\right]_x \!= \frac{d_x^{(\hn\cdot\hn)}}{\lambda}\sin\vartheta^{(\hn\cdot\hn)}, \label{eq:naf_ell}\\
\eta^{(\hn\cdot\hn)}  = \frac{d_z^{(\hn\cdot\hn)}}{\lambda}\,\left[\mathbf{u}^{(\hn\cdot\hn)}\right]_z \!= \frac{d_z^{(\hn\cdot\hn)}}{\lambda}\sin\varphi^{(\hn\cdot\hn)}. \label{eq:naf_eta}
\end{align}
By utilizing conic angles for our \gls{naf} definitions, the \gls{naf} scalar
projections are decoupled by construction, as~\eqref{eq:ElementPhase} shows. This contrasts with the conventional azimuth $\theta$ and elevation $\varphi$ of the literature~\cite{treesOptimumArrayProcessing2002}, where $\mathbf{u}=[\cos\varphi\sin\theta,\cos\varphi\cos\theta,\sin\varphi]^\intercal$. For a planar array, changes to one \gls{naf} component do not affect the other, so the angular sampling task can be solved independently per component and combined via a Kronecker product~\cite{mandelli2022sampling}. 
In this bistatic scenario, the azimuth components of \gls{tx} and \gls{rx} are likewise mutually decoupled, and their joint azimuth sampling task is solved in~\cite{felixOptimalAzimuthSampling2025a} as the Cartesian product of the individual azimuth sets. By \gls{dft} sampling theory, the per-array minimal azimuth \gls{naf} sampling set is $\mathcal{U}^{(\hn\cdot\hn)}\subset[-\frac{1}{2},\frac{1}{2})$ of size $N_x^{(\hn\cdot\hn)}$, uniformly spaced. The full bistatic azimuth set is then $\mathcal{P} = \mathcal{U}^{(\mathrm{t})} \times \mathcal{U}^{(\mathrm{r})}$. Each array's azimuth response is reconstructed from its $N_x^{(\hn\cdot\hn)}$ uniform \gls{naf} samples by Dirichlet kernel interpolation~\cite{mandelli2022sampling, treesOptimumArrayProcessing2002}. 

The elevation domain, by contrast, does not decouple.
For joint bistatic processing the spatially separated \gls{tx} and \gls{rx} must illuminate a common \textit{scan point}, as in Fig.~\ref{fig:bistatic_ang_pairs}. This \textit{scan-on-scan} constraint~\cite{willisBistaticRadar2005} couples the elevation angles $\varphi^{(\mathrm{t})}$ and $\varphi^{(\mathrm{r})}$ for any given azimuth pair $(\vartheta^{(\mathrm{t})},\vartheta^{(\mathrm{r})})$.
Analog, hybrid, and even fully digital beamforming acquire only a limited subset of the direction pairs within $\hat{a}(\mathbf{u}^{(\mathrm{t})},\mathbf{u}^{(\mathrm{r})})$ at a time, since the \gls{tx} must beamform toward one direction, motivating the sampling design that follows.

\section{Bistatic Angular Sampling Task} \label{sc:proposal}

Contrary to the azimuth task of~\ref{sc:model}, the elevation task does not decouple into per-array uniform \gls{naf} sampling. We derive the scan-on-scan elevation coupling and cast the joint elevation aperture of the bistatic pair as an ortho-baseline coarray. This allows us to obtain the minimal sample set for the full bistatic angular domain, with near-lossless reconstruction whose residual error is bounded in Appendix~\ref{ap:agcd}.

\subsection{Scan-on-Scan Elevation Coupling} \label{sc:proposal:coupling}

For a fixed azimuth pair $\left(\vartheta^{(\mathrm{t})}, \vartheta^{(\mathrm{r})}\right)$ the azimuth angular cone surfaces intersect on a circumference of radius $r_c$ centered at $x^{\star}$ on the baseline, as illustrated in Fig.~\ref{fig:bistatic_ang_pairs}. Solving the cone-intersection geometry yields $r_c^2 = b^2/\left(\tan\vartheta^{(\mathrm{t})} - \tan\vartheta^{(\mathrm{r})}\right)^2$ and $x^{\star} = b\,\tan\vartheta^{(\mathrm{t})}/\left(\tan\vartheta^{(\mathrm{t})} - \tan\vartheta^{(\mathrm{r})}\right)$. Any scan point on this circumference has constant slant ranges $r^{(\mathrm{t})} = \sqrt{(x^{\star})^2 + r_c^2}$ and $r^{(\mathrm{r})} = \sqrt{(x^{\star}-b)^2 + r_c^2}$.
For a scan point at height $z_c$ on this circumference, the conic elevation angle for a scan direction $\mathbf{u}^{(\hn\cdot\hn)}$ is $\sin\varphi^{(\hn\cdot\hn)}=\left[\mathbf{u}^{(\hn\cdot\hn)}\right]_z=z_c/r^{(\hn\cdot\hn)}$, so for each array
\begin{equation}
    \sin\varphi^{(\mathrm{t})}=\frac{z_c}{r^{(\mathrm{t})}}, \qquad
\sin\varphi^{(\mathrm{r})}=\frac{z_c}{r^{(\mathrm{r})}}.
    \label{eq:conicelevationangles}
\end{equation}
Since the scan point is common to both arrays (the scan-on-scan constraint of~\ref{sc:model}), we get
\begin{equation}
    \sin\varphi^{(\mathrm{t})}=\frac{r^{(\mathrm{r})}}{r^{(\mathrm{t})}}\sin\varphi^{(\mathrm{r})}=k^\star\!\left(\vartheta^{(\mathrm{t})}, \vartheta^{(\mathrm{r})}\right)\sin\varphi^{(\mathrm{r})},
    \label{eq:conicelevationequal}
\end{equation}
where the \textit{bistatic angular translation factor} $k^\star = r^{(\mathrm{r})}/r^{(\mathrm{t})}$ maps the conic elevation angle of one array to the other. Since $\cos\vartheta^{(\hn\cdot\hn)} = r_c/r^{(\hn\cdot\hn)}$ on the circumference, the baseline cancels and $k^\star = \cos\vartheta^{(\mathrm{t})}/\cos\vartheta^{(\mathrm{r})}$ depends on the azimuth pair alone, so the elevation sampling below is invariant to the baseline length $b$. The factor grows without bound only toward \gls{rx} endfire, $\cos\vartheta^{(\mathrm{r})} \to 0$, where it is capped in~\ref{sc:proposal:coarray}. From~\eqref{eq:conicelevationequal}, the \gls{tx} elevation \gls{naf} component follows as
\begin{equation}
    \eta^{(\mathrm{t})} = \frac{d_z^{(\mathrm{t})}}{d_z^{(\mathrm{r})}} \, k^\star \, \eta^{(\mathrm{r})}.
    \label{eq:elevationnaf}
\end{equation}

\subsection{Ortho-Baseline Coarray} \label{sc:proposal:coarray}

Substituting the scan-on-scan relation~\eqref{eq:conicelevationequal} into the steering phase~\eqref{eq:BeamformedSignalBistaticRx}, the combined bistatic phase argument of the element pair $\left(n,m\right)$ (included in the exponent as $e^{-jk_0\Theta_{n,m}}$) becomes
\begin{multline}
\Theta_{n,m}\left(\vartheta^{(\mathrm{t})}\!, \vartheta^{(\mathrm{r})}\!, \varphi^{(\mathrm{r})}\right) = \mathbf{u}^{(\mathrm{t})\intercal}\mathbf{p}_n^{(\mathrm{t})}
  + \mathbf{u}^{(\mathrm{r})\intercal}\mathbf{p}_m^{(\mathrm{r})} \\
= x_n^{(\mathrm{t})}\sin\vartheta^{(\mathrm{t})}
  + x_m^{(\mathrm{r})}\sin\vartheta^{(\mathrm{r})}
  + c_{n,m}\sin\varphi^{(\mathrm{r})},
\label{eq:CombinedPhaseCoarray}
\end{multline}
where $c_{n,m} = k^\star z_n^{(\mathrm{t})} + z_m^{(\mathrm{r})}$ is a virtual ortho-baseline ($z$) coordinate.
The $k^{\star}$ factor maps the \gls{tx} elevation onto the \gls{rx} axis, so $c_{n,m}$ pairs with $\sin\varphi^{(\mathrm{r})}$ in~\eqref{eq:CombinedPhaseCoarray} 
exactly as a physical element at height $c_{n,m}$ would. These virtual elements form the \textit{ortho-baseline coarray}
\begin{equation}
\mathcal{C} = \left\{ k^{\star} z^{(\mathrm{t})} + z^{(\mathrm{r})} \;\middle|\; z^{(\mathrm{t})} \in \mathcal{Z}^{(\mathrm{t})},\; z^{(\mathrm{r})} \in \mathcal{Z}^{(\mathrm{r})} \right\}\!,
\label{eq:orthocoarray}
\end{equation}
with $\mathcal{Z}^{(\hn\cdot\hn)}$ the per-array ortho-baseline element coordinates. It is the bistatic counterpart of the monostatic sum coarray~\cite{hoctor1990unifying}.

After substitution, the elevation response for a fixed azimuth pair depends only on the offset between scan and target elevation \gls{naf} coordinates. This makes it shift-invariant in $\eta^{(\mathrm{r})}$, enabling uniform \gls{naf} elevation treatment.

The azimuth scheme of~\ref{sc:model} reconstructs each array from uniform \gls{naf} samples~\cite{felixOptimalAzimuthSampling2025a}. Reusing this scheme in elevation requires the ortho-baseline coarray $\mathcal{C}$ to lie on a regular grid. Since $k^\star$ is generally irrational, the $k^{\star}$-scaled \gls{tx} spacing $k^{\star} d_z^{(\mathrm{t})}$ and the \gls{rx} spacing $d_z^{(\mathrm{r})}$ are incommensurate, so $\mathcal{C}$ is non-uniform. We therefore replace the ill-conditioned exact divisor with an \gls{agcd} $\tilde{d}_{\mathcal{C}} \approx \gcd\!\left(k^{\star} d_z^{(\mathrm{t})}/\lambda,\; d_z^{(\mathrm{r})}/\lambda\right)$, the coarsest spacing that divides both arguments to within a tolerance $\kappa$ in \gls{naf} units, with the construction detailed in Appendix~\ref{ap:agcd}. The same tolerance caps the translation factor at $k^{\star}_{\max} = 1/(2\kappa)$ toward \gls{rx} endfire, keeping the aperture $L$ and the order $K$ finite. The \gls{agcd} spacing restores a regular grid of \gls{naf} aperture $L = \left[k^{\star}(N_z^{(\mathrm{t})}-1)\,d_z^{(\mathrm{t})} + (N_z^{(\mathrm{r})}-1)\,d_z^{(\mathrm{r})}\right]/\lambda$, the $k^{\star}$-scaled \gls{tx} ortho-baseline aperture plus the \gls{rx} ortho-baseline aperture, on which the uniform-sampling reconstruction applies. For such an aperture, the elevation response is band-limited to order $K = \lceil L\,\lambda/d_z^{(\mathrm{r})} \rceil + 1$ on the \gls{rx} elevation-\gls{naf} axis.

Aggregating the \gls{tx} and \gls{rx} ortho-baseline apertures into a single virtual array yields the joint bistatic elevation resolution on a single \gls{naf} axis, the basis for the minimal elevation sampling this work develops.

\subsection{Sampling and Reconstruction} \label{sc:proposal:sampling}

For a fixed azimuth pair, the bistatic angular sampling task selects $K$ uniform \gls{rx} elevation-\gls{naf} samples according to Nyquist--Shannon.
With the \gls{tx} values linked through~\eqref{eq:elevationnaf}, these form the elevation set $\mathcal{U}_z\!\left(\vartheta^{(\mathrm{t})},\vartheta^{(\mathrm{r})}\right)$ of $K$ coupled pairs $\left(\eta^{(\mathrm{t})},\eta^{(\mathrm{r})}\right)$.
Collecting the elevation sets across all azimuth pairs yields the full \gls{4D} bistatic angular sampling set
\begin{equation}
\mathcal{B} = \bigcup_{(\vartheta^{(\mathrm{t})},\vartheta^{(\mathrm{r})}) \, \in \, \mathcal{P}}
\left\{\left(\vartheta^{(\mathrm{t})},\vartheta^{(\mathrm{r})}\right)\right\}
\times \mathcal{U}_z\!\left(\vartheta^{(\mathrm{t})},\vartheta^{(\mathrm{r})}\right),
\label{eq:bistaticsamplingset}
\end{equation}
where $\mathcal{P}$ is the azimuth set of~\ref{sc:model}. As such, the elevation sampling extends, but does not redefine, the azimuth sampling task of~\ref{sc:model}.

The $K$ samples of each elevation set $\mathcal{U}_z$ are uniform over the \gls{rx} elevation-\gls{naf} period $\eta^{(\mathrm{r})}\in[-\tfrac{1}{2},\tfrac{1}{2})$, of which only the visible region $\eta^{(\mathrm{r})}\in[-d_z^{(\mathrm{r})}/\lambda,\, d_z^{(\mathrm{r})}/\lambda)$ maps to physical elevation directions. A practical system acquires only this physical subset together with a guard margin of $\nu_{\mathrm{guard}}$ samples (one or two) beyond each boundary, which preserves reconstruction quality near the visible-region edge, and discards the remaining non-physical samples.

The minimal set $\mathcal{B}$ fully determines the band-limited angular response, so any direction not visited by the sampling task can be recovered by Dirichlet kernel interpolation~\cite{mandelli2022sampling, treesOptimumArrayProcessing2002}, the periodic finite-aperture form of the Whittaker--Shannon theorem. On a uniform sampling lattice, the Dirichlet kernel
\begin{equation}
    D_K(u) = \frac{\sin(\pi K u)}{K \sin(\pi u)}
    \label{eq:dirichlet}
\end{equation}
achieves perfect noiseless reconstruction, where $u$ is a \gls{naf} coordinate in which the response has unit period. The azimuth axes reconstruct with~\eqref{eq:dirichlet} at $u = \ell^{(\hn\cdot\hn)}$ and the per-array order $N_x^{(\hn\cdot\hn)}$, following the azimuth scheme of~\ref{sc:model}. The $K$ coupled elevation samples of each ortho-baseline coarray reconstruct the same way, with~\eqref{eq:dirichlet} at $u = \eta^{(\mathrm{r})}$ and the elevation order $K$, treated as a uniform elevation-\gls{naf} axis. This direct band-limited form is the reconstruction evaluated in~\ref{sc:simulation}. The coarray elements actually lie on the finer \gls{agcd} grid, and matching the kernel period to it resolves more structure, but inferring that higher order from the $K$ acquired samples extrapolates beyond the critically sampled bandwidth and amplifies noise, making the direct form the noise-robust choice.

With the elevation and azimuth kernels, the \gls{3D} image is reconstructed elevation-first, as summarized in Algorithm~\ref{alg:basiis}. Azimuth-first interpolation was discarded, as each azimuth pair carries a different order $K$ and spacing $\tilde{d}_{\mathcal{C}}$, so the elevation axes must first share a common $\eta$ grid of $K_{\max}$ points (the largest $K$). The minimal set $\mathcal{B}$ enables near-lossless reconstruction of the complete \gls{3D} bistatic angular image from the fewest acquired \gls{tx}--\gls{rx} direction pairs, with the residual error of the coarray grid set by the \gls{agcd} tolerance $\kappa$ and bounded in Appendix~\ref{ap:agcd}.

\begin{algorithm}[t]
\caption{BASIIS acquisition and reconstruction}
\label{alg:basiis}
\begin{algorithmic}[1]
\State \textbf{acquire} $\hat{a}$ over the minimal set $\mathcal{B}$~\eqref{eq:bistaticsamplingset}
\ForAll{$(\vartheta^{(\mathrm{t})},\vartheta^{(\mathrm{r})}) \in \mathcal{P}$}
    \State interpolate the $K$ elevation samples with $D_K$~\eqref{eq:dirichlet}
    \State resample onto the common $\eta$ grid of $K_{\max}$ points
\EndFor
\ForAll{cells of the common $\eta$ grid}
    \State interpolate \gls{tx} and \gls{rx} azimuth with $D_{N_x^{(\hn\cdot\hn)}}$
\EndFor
\State \textbf{return} \gls{3D} image over $\left(\ell^{(\mathrm{t})}\!, \ell^{(\mathrm{r})}\!, \eta^{(\mathrm{r})}\right)$
\end{algorithmic}
\end{algorithm}
The acquisition count and the reconstruction cost follow from this structure. The acquisition comprises \mbox{$|\mathcal{B}| = \sum_{(\vartheta^{(\mathrm{t})},\vartheta^{(\mathrm{r})})\in\mathcal{P}} K(\vartheta^{(\mathrm{t})},\vartheta^{(\mathrm{r})}) \le N_x^{(\mathrm{t})} N_x^{(\mathrm{r})} K_{\max}$} \gls{tx}--\gls{rx} direction pairs: the $N_x^{(\mathrm{t})} N_x^{(\mathrm{r})}$ azimuth pairs of $\mathcal{P}$, each carrying $K$ coupled elevation samples. Both the per-array azimuth count $N_x^{(\hn\cdot\hn)}$ and the elevation order $K$ are the Nyquist--Shannon-minimal counts for their respective apertures, so $\mathcal{B}$ is the smallest per-pair uniform set of this form. Reconstruction proceeds elevation-first, with each \gls{dft}-based Dirichlet interpolation costing $O(\hn K_{\max}\log\hn K_{\max}\hn)$, for a total cost of $O(G\log G)$ in the number $G \ge |\mathcal{B}|$ of reconstructed image cells, set by the \acrfull{if} of~\ref{sc:simulation}.

\section{Numerical Evaluation}
\label{sc:simulation}

\renewcommand{\arraystretch}{1.15} %
\begin{table}
    \centering
    \caption{Simulation parameters and assumptions}
    \label{tab:sim_param}
    \begin{tabular}{|c|c|}
        \hline
        \textbf{Parameter} & \textbf{Value / Description} \\
        \Xhline{3\arrayrulewidth}
        TX/RX array & \glspl{ura}, $N_x^{(\hn\cdot\hn)} = N_z^{(\hn\cdot\hn)} = 11$ \\
        \hline
        TX/RX element spacing & $d_x^{(\hn\cdot\hn)} = d_z^{(\hn\cdot\hn)} = \lambda/2$ \\
        \hline
        Bistatic baseline $b$ & \SI{10}{m} \\
        \hline
        Number of scatterers & $2$ \\
        \hline
        Image-plane SNR & noiseless or $\approx \SI{17.6}{dB}$ ($\sigma_N^2 = \SI{38}{dB}$) \\
        \hline
        Scatterer reflection coefficients & $e^{j\Phi_s}$, $\Phi_s \sim U[0,2\pi)$ \\
        \hline
        \gls{agcd} tolerance $\kappa$ & $10^{-3}$ \\
        \hline
        BASIIS acquisition margin & $(\nu_{\mathrm{add}},\,\nu_{\mathrm{guard}})=(2,2)$ \\
        \hline
        Desired $P^{(\mathrm{FA})}$ of CFAR & $10^{-9}$ \\
        \hline
        Iterations per data point & $10^{4}$ \\
        \hline
    \end{tabular}
\end{table}

The bistatic setup follows the parameters in Table~\ref{tab:sim_param}.
The two scatterers are confined to the \gls{naf} range $[-0.4, 0.4]$, equivalent to conic angles within $\pm 53.13^\circ$ of broadside, keeping them clear of the visible-region edges at $\pm 0.5$. Near these edges a method-independent false-alarm spike, intrinsic to the \gls{naf} visible-region boundary, would otherwise confound the comparison. Both arrays carry a \SI{45}{dB} Chebyshev taper~\cite{Lynch1997Chebyshev} along the baseline and a uniform taper along the ortho-baseline.

The proposed BASIIS acquisition uses the AGCD-aligned ortho-baseline coarray of~\ref{sc:proposal:coarray}, sampled in the visible region as prescribed in~\ref{sc:proposal:sampling}, with reconstruction by the direct band-limited Dirichlet kernel~\eqref{eq:dirichlet} at the elevation order $K$. 
The BASIIS approach carries two acquisition margins, $\nu_{\mathrm{add}}$ extra per-pair Nyquist samples and the visible-region guard $\nu_{\mathrm{guard}}$. We evaluate at the operating point with $|\mathcal{B}| = 3441$ direction pairs over $121$ azimuth pairs, each carrying $K \in [16, K_{\max}]$ elevation samples with $K_{\max} = 36$, on average $22$, plus the $\nu_{\mathrm{add}} + 2\nu_{\mathrm{guard}} = 6$ sample margin. The spread of $K$ reflects $k^{\star}$ over the $121$ azimuth pairs, the only quantity through which the geometry enters (\ref{sc:proposal:coupling}), so $|\mathcal{B}|$, the \gls{agcd} grids and the savings below hold for any baseline length. Only rotated arrays, excluded in~\ref{sc:model}, would alter them. The tolerance $\kappa = 10^{-3}$ trades reconstruction fidelity against the elevation order $K$, capping $k^{\star}$ at $k^{\star}_{\max} = 500$ while retaining near-lossless reconstruction.

As references we use oversampled imaging at three \glspl{if} $\in \{1.25,\,1.36,\,2.00\}$ in two variants. The \emph{isotropic} reference oversamples all three angular axes by the \gls{if}, acquiring $\lceil \mathrm{IF}\cdot N_x\rceil^3 = 2744/3375/10648$ direction pairs. The \emph{anisotropic} reference oversamples only the azimuth axes and fixes the elevation axis at the BASIIS reconstruction-grid density $K_{\max} + \nu_{\mathrm{add}} = 38$ ($7448/8550/18392$ pairs), equalizing the elevation straddle loss, the attenuation of a target falling between image cells. The isotropic case is an independent dense reference, while the anisotropic one is a BASIIS-informed control isolating the cell-density mechanism.
For BASIIS the \gls{if} carries no acquisition cost. It only sets the reconstruction grid: each axis is interpolated to the reference density where the native grid is coarser (azimuth axes) and left unchanged where it is already denser (the elevation axis at $K_{\max} + \nu_{\mathrm{add}}$). The acquisition stays at $|\mathcal{B}| = 3441$ for every \gls{if}, whereas for the oversampled references a higher \gls{if} directly multiplies the acquired direction pairs.

\begin{figure*}[ht]
    \def\scale{.595}
    \def\subfigwidth{0.329}
    \def\genvspace{-1.6mm}
    \def\genhspace{-5.0mm}
    \def\figwidth{9.0cm}
    \def\figheight{5.5cm}
    \def\xmin{0.08}
    \def\xmax{0.28}
    \centering
    \centerline{{\ref{legend3way}}}
    \begin{subfigure}[t]{\subfigwidth\textwidth}\centering\begin{tikzpicture}
\begin{semilogyaxis}[
    xlabel={Target NAF separation $\Delta\ell^{(\mathrm{r})}$},
    ylabel={$P^{(\mathrm{MD})}$},
    xlabel style={font=\footnotesize},
    ylabel style={font=\footnotesize},
    y label style={yshift=-1.5em},
    ytick={0.001, 0.002, 0.003, 0.004, 0.005, 0.006, 0.007, 0.008, 0.009, 0.01, 0.02, 0.03, 0.04, 0.05, 0.06, 0.07, 0.08, 0.09, 0.1, 0.2, 0.3, 0.4, 0.5, 0.6, 0.7, 0.8, 0.9, 1},
    yticklabels={$10^{-3}$, , , , , , , , , , , , , , , , , ,   , , , , , , , , , $10^{0}$},
    ticklabel style={font=\footnotesize},
    xmin=\xmin, xmax=\xmax,
    ymin=1e-3, ymax=1e0,  %
    grid=both,
    grid style={solid, black!15},
    every axis plot/.append style={very thick},
    legend cell align={left},
    legend columns=-1,
    legend style={font=\scriptsize, draw=lightgray},
    legend to name=legend3way,
    width=\figwidth, height=\figheight, scale=\scale
]
\addlegendimage{white, fill=white}\addlegendentry{\hspace{-.7cm}IF:\;\;\;}
\addlegendimage{very thick, teal,   solid}\addlegendentry{$1.25$}
\addlegendimage{very thick, violet, solid}\addlegendentry{$1.36$}
\addlegendimage{very thick, orange, solid}\addlegendentry{$2.00$}
\addlegendimage{white, fill=white}\addlegendentry{Method:\;\;\;}
\addlegendimage{very thick, black, solid}\addlegendentry{BASIIS}
\addlegendimage{very thick, black, dotted}\addlegendentry{Oversampled isotropic}
\addlegendimage{very thick, black, dashed}\addlegendentry{Oversampled anisotropic}
\addplot[very thick, teal,   solid, forget plot] table[x index=0, y index=1] {Data/3D/scenario_A/prob_missed_detect_3way.txt};
\addplot[very thick, violet, solid, forget plot] table[x index=0, y index=2] {Data/3D/scenario_A/prob_missed_detect_3way.txt};
\addplot[very thick, orange, solid, forget plot] table[x index=0, y index=3] {Data/3D/scenario_A/prob_missed_detect_3way.txt};
\addplot[very thick, teal,   dotted, forget plot] table[x index=0, y index=4] {Data/3D/scenario_A/prob_missed_detect_3way.txt};
\addplot[very thick, violet, dotted, forget plot] table[x index=0, y index=5] {Data/3D/scenario_A/prob_missed_detect_3way.txt};
\addplot[very thick, orange, dotted, forget plot] table[x index=0, y index=6] {Data/3D/scenario_A/prob_missed_detect_3way.txt};
\addplot[very thick, teal,   dashed, forget plot] table[x index=0, y index=7] {Data/3D/scenario_A/prob_missed_detect_3way.txt};
\addplot[very thick, violet, dashed, forget plot] table[x index=0, y index=8] {Data/3D/scenario_A/prob_missed_detect_3way.txt};
\addplot[very thick, orange, dashed, forget plot] table[x index=0, y index=9] {Data/3D/scenario_A/prob_missed_detect_3way.txt};
\end{semilogyaxis}
\end{tikzpicture}\vspace*{\genvspace}\caption{$P^{(\mathrm{MD})}$, noiseless.}\label{fig:3d_pmd_3way_noiseless_m}\end{subfigure}\hspace*{\genhspace}%
    \begin{subfigure}[t]{\subfigwidth\textwidth}\centering\begin{tikzpicture}
\begin{semilogyaxis}[
    xlabel={Target NAF separation $\Delta\ell^{(\mathrm{r})}$},
    ylabel={\gls{naf} RMSE},
    xlabel style={font=\footnotesize},
    ylabel style={font=\footnotesize},
    ylabel style={yshift=-1.5em},
    ytick={0.00001, 0.00002, 0.00003, 0.00004, 0.00005, 0.00006, 0.00007, 0.00008, 0.00009, 0.0001, 0.0002, 0.0003, 0.0004, 0.0005, 0.0006, 0.0007, 0.0008, 0.0009, 0.001, 0.002, 0.003, 0.004, 0.005, 0.006, 0.007, 0.008, 0.009, 0.01, 0.02, 0.03, 0.04, 0.05, 0.06, 0.07, 0.08, 0.09, 0.1, 0.2, 0.3, 0.4, 0.5, 0.6, 0.7, 0.8, 0.9, 1},
    yticklabels={$10^{-5}$, , , , , , , , , $10^{-4}$, , , , , , , , , $10^{-3}$, , , , , , , , , $10^{-2}$, , , , , , , , ,   , , , , , , , , , $10^{0}$},
    ticklabel style={font=\footnotesize},
    xmin=\xmin, xmax=\xmax,
    ymin=1e-2, ymax=1e0,
    grid=both,
    grid style={solid, black!15},
    every axis plot/.append style={very thick},
    width=\figwidth, height=\figheight, scale=\scale
]
\addplot[very thick, teal,   solid] table[x index=0, y index=1] {Data/3D/scenario_A/rmse_3way.txt};
\addplot[very thick, violet, solid] table[x index=0, y index=2] {Data/3D/scenario_A/rmse_3way.txt};
\addplot[very thick, orange, solid] table[x index=0, y index=3] {Data/3D/scenario_A/rmse_3way.txt};
\addplot[very thick, teal,   dotted] table[x index=0, y index=4] {Data/3D/scenario_A/rmse_3way.txt};
\addplot[very thick, violet, dotted] table[x index=0, y index=5] {Data/3D/scenario_A/rmse_3way.txt};
\addplot[very thick, orange, dotted] table[x index=0, y index=6] {Data/3D/scenario_A/rmse_3way.txt};
\addplot[very thick, teal,   dashed] table[x index=0, y index=7] {Data/3D/scenario_A/rmse_3way.txt};
\addplot[very thick, violet, dashed] table[x index=0, y index=8] {Data/3D/scenario_A/rmse_3way.txt};
\addplot[very thick, orange, dashed] table[x index=0, y index=9] {Data/3D/scenario_A/rmse_3way.txt};
\end{semilogyaxis}
\end{tikzpicture}\vspace*{\genvspace}\caption{\gls{naf} RMSE, noiseless.}\label{fig:3d_rmse_3way_noiseless_m}\end{subfigure}\hspace*{\genhspace}%
    \begin{subfigure}[t]{\subfigwidth\textwidth}\centering\begin{tikzpicture}
\begin{axis}[
    xlabel={Target NAF separation $\Delta\ell^{(\mathrm{r})}$},
    ylabel={$F_1$ score},
    xlabel style={font=\footnotesize},
    ylabel style={font=\footnotesize},
    ylabel style={yshift=-1.2em},
    ytick={0.4, 0.5, 0.6, 0.7, 0.8, 0.9, 1},
    yticklabels={0.4, 0.5, 0.6,   ,  ,  , 1},
    ticklabel style={font=\footnotesize},
    xmin=\xmin, xmax=\xmax,
    ymin=0.58, ymax=1.02,
    grid=both,
    grid style={solid, black!15},
    every axis plot/.append style={very thick},
    width=\figwidth, height=\figheight, scale=\scale
]
\addplot[very thick, teal,   solid] table[x index=0, y index=1] {Data/3D/scenario_A/f1_3way.txt};
\addplot[very thick, violet, solid] table[x index=0, y index=2] {Data/3D/scenario_A/f1_3way.txt};
\addplot[very thick, orange, solid] table[x index=0, y index=3] {Data/3D/scenario_A/f1_3way.txt};
\addplot[very thick, teal,   dotted] table[x index=0, y index=4] {Data/3D/scenario_A/f1_3way.txt};
\addplot[very thick, violet, dotted] table[x index=0, y index=5] {Data/3D/scenario_A/f1_3way.txt};
\addplot[very thick, orange, dotted] table[x index=0, y index=6] {Data/3D/scenario_A/f1_3way.txt};
\addplot[very thick, teal,   dashed] table[x index=0, y index=7] {Data/3D/scenario_A/f1_3way.txt};
\addplot[very thick, violet, dashed] table[x index=0, y index=8] {Data/3D/scenario_A/f1_3way.txt};
\addplot[very thick, orange, dashed] table[x index=0, y index=9] {Data/3D/scenario_A/f1_3way.txt};
\end{axis}
\end{tikzpicture}\vspace*{\genvspace}\caption{$F_1$ score, noiseless.}\label{fig:3d_f1_3way_noiseless_m}\end{subfigure}\\[1mm]
    \begin{subfigure}[t]{\subfigwidth\textwidth}\centering\begin{tikzpicture}
\begin{semilogyaxis}[
    xlabel={Target NAF separation $\Delta\ell^{(\mathrm{r})}$},
    ylabel={$P^{(\mathrm{MD})}$},
    xlabel style={font=\footnotesize},
    ylabel style={font=\footnotesize},
    y label style={yshift=-1.5em},
    ytick={0.001, 0.002, 0.003, 0.004, 0.005, 0.006, 0.007, 0.008, 0.009, 0.01, 0.02, 0.03, 0.04, 0.05, 0.06, 0.07, 0.08, 0.09, 0.1, 0.2, 0.3, 0.4, 0.5, 0.6, 0.7, 0.8, 0.9, 1},
    yticklabels={$10^{-3}$, , , , , , , , , , , , , , , , , ,   , , , , , , , , , $10^{0}$},
    ticklabel style={font=\footnotesize},
    xmin=\xmin, xmax=\xmax,
    ymin=1e-3, ymax=1e0,  %
    grid=both,
    grid style={solid, black!15},
    every axis plot/.append style={very thick},
    width=\figwidth, height=\figheight, scale=\scale
]
\addplot[very thick, teal,   solid] table[x index=0, y index=10] {Data/3D/scenario_A/prob_missed_detect_3way.txt};
\addplot[very thick, violet, solid] table[x index=0, y index=11] {Data/3D/scenario_A/prob_missed_detect_3way.txt};
\addplot[very thick, orange, solid] table[x index=0, y index=12] {Data/3D/scenario_A/prob_missed_detect_3way.txt};
\addplot[very thick, teal,   dotted] table[x index=0, y index=13] {Data/3D/scenario_A/prob_missed_detect_3way.txt};
\addplot[very thick, violet, dotted] table[x index=0, y index=14] {Data/3D/scenario_A/prob_missed_detect_3way.txt};
\addplot[very thick, orange, dotted] table[x index=0, y index=15] {Data/3D/scenario_A/prob_missed_detect_3way.txt};
\addplot[very thick, teal,   dashed] table[x index=0, y index=16] {Data/3D/scenario_A/prob_missed_detect_3way.txt};
\addplot[very thick, violet, dashed] table[x index=0, y index=17] {Data/3D/scenario_A/prob_missed_detect_3way.txt};
\addplot[very thick, orange, dashed] table[x index=0, y index=18] {Data/3D/scenario_A/prob_missed_detect_3way.txt};
\end{semilogyaxis}
\end{tikzpicture}\vspace*{\genvspace}\caption{$P^{(\mathrm{MD})}$, $\approx\SI{17.6}{dB}$.}\label{fig:3d_pmd_3way_noise38_m}\end{subfigure}\hspace*{\genhspace}%
    \begin{subfigure}[t]{\subfigwidth\textwidth}\centering\begin{tikzpicture}
\begin{semilogyaxis}[
    xlabel={Target NAF separation $\Delta\ell^{(\mathrm{r})}$},
    ylabel={\gls{naf} RMSE},
    xlabel style={font=\footnotesize},
    ylabel style={font=\footnotesize},
    ylabel style={yshift=-1.5em},
    ytick={0.00001, 0.00002, 0.00003, 0.00004, 0.00005, 0.00006, 0.00007, 0.00008, 0.00009, 0.0001, 0.0002, 0.0003, 0.0004, 0.0005, 0.0006, 0.0007, 0.0008, 0.0009, 0.001, 0.002, 0.003, 0.004, 0.005, 0.006, 0.007, 0.008, 0.009, 0.01, 0.02, 0.03, 0.04, 0.05, 0.06, 0.07, 0.08, 0.09, 0.1, 0.2, 0.3, 0.4, 0.5, 0.6, 0.7, 0.8, 0.9, 1},
    yticklabels={$10^{-5}$, , , , , , , , , $10^{-4}$, , , , , , , , , $10^{-3}$, , , , , , , , , $10^{-2}$, , , , , , , , ,   , , , , , , , , , $10^{0}$},
    ticklabel style={font=\footnotesize},
    xmin=\xmin, xmax=\xmax,
    ymin=1e-2, ymax=1e0,
    grid=both,
    grid style={solid, black!15},
    every axis plot/.append style={very thick},
    width=\figwidth, height=\figheight, scale=\scale
]
\addplot[very thick, teal,   solid] table[x index=0, y index=10] {Data/3D/scenario_A/rmse_3way.txt};
\addplot[very thick, violet, solid] table[x index=0, y index=11] {Data/3D/scenario_A/rmse_3way.txt};
\addplot[very thick, orange, solid] table[x index=0, y index=12] {Data/3D/scenario_A/rmse_3way.txt};
\addplot[very thick, teal,   dotted] table[x index=0, y index=13] {Data/3D/scenario_A/rmse_3way.txt};
\addplot[very thick, violet, dotted] table[x index=0, y index=14] {Data/3D/scenario_A/rmse_3way.txt};
\addplot[very thick, orange, dotted] table[x index=0, y index=15] {Data/3D/scenario_A/rmse_3way.txt};
\addplot[very thick, teal,   dashed] table[x index=0, y index=16] {Data/3D/scenario_A/rmse_3way.txt};
\addplot[very thick, violet, dashed] table[x index=0, y index=17] {Data/3D/scenario_A/rmse_3way.txt};
\addplot[very thick, orange, dashed] table[x index=0, y index=18] {Data/3D/scenario_A/rmse_3way.txt};
\end{semilogyaxis}
\end{tikzpicture}\vspace*{\genvspace}\caption{\gls{naf} RMSE, $\approx\SI{17.6}{dB}$.}\label{fig:3d_rmse_3way_noise38_m}\end{subfigure}\hspace*{\genhspace}%
    \begin{subfigure}[t]{\subfigwidth\textwidth}\centering\begin{tikzpicture}
\begin{axis}[
    xlabel={Target NAF separation $\Delta\ell^{(\mathrm{r})}$},
    ylabel={$F_1$ score},
    xlabel style={font=\footnotesize},
    ylabel style={font=\footnotesize},
    ylabel style={yshift=-1.2em},
    ytick={0.4, 0.5, 0.6, 0.7, 0.8, 0.9, 1},
    yticklabels={0.4, 0.5, 0.6,   ,   ,   , 1},
    ticklabel style={font=\footnotesize},
    xmin=\xmin, xmax=\xmax,
    ymin=0.58, ymax=1.02,
    grid=both,
    grid style={solid, black!15},
    every axis plot/.append style={very thick},
    width=\figwidth, height=\figheight, scale=\scale
]
\addplot[very thick, teal,   solid] table[x index=0, y index=10] {Data/3D/scenario_A/f1_3way.txt};
\addplot[very thick, violet, solid] table[x index=0, y index=11] {Data/3D/scenario_A/f1_3way.txt};
\addplot[very thick, orange, solid] table[x index=0, y index=12] {Data/3D/scenario_A/f1_3way.txt};
\addplot[very thick, teal,   dotted] table[x index=0, y index=13] {Data/3D/scenario_A/f1_3way.txt};
\addplot[very thick, violet, dotted] table[x index=0, y index=14] {Data/3D/scenario_A/f1_3way.txt};
\addplot[very thick, orange, dotted] table[x index=0, y index=15] {Data/3D/scenario_A/f1_3way.txt};
\addplot[very thick, teal,   dashed] table[x index=0, y index=16] {Data/3D/scenario_A/f1_3way.txt};
\addplot[very thick, violet, dashed] table[x index=0, y index=17] {Data/3D/scenario_A/f1_3way.txt};
\addplot[very thick, orange, dashed] table[x index=0, y index=18] {Data/3D/scenario_A/f1_3way.txt};
\end{axis}
\end{tikzpicture}\vspace*{\genvspace}\caption{$F_1$ score, $\approx\SI{17.6}{dB}$.}\label{fig:3d_f1_3way_noise38_m}\end{subfigure}
    \caption{Detection and localization over the \gls{rx} \gls{naf} separation $\Delta\ell^{(\mathrm{r})}$ of two scatterers, for BASIIS (solid) against isotropic (dotted) and anisotropic (dashed) oversampling at IF $\in \{1.25,\,1.36,\,2.00\}$. Noiseless (top row), all methods reach parity, so acquisition cost differentiates. At image-plane SNR $\approx\SI{17.6}{dB}$ (bottom row), BASIIS tracks the anisotropic reference and leads the isotropic one at lower IF.}
    \label{fig:3d_eval_3way_merged}
    \vspace{-6mm}
\end{figure*}

\paragraph{Detector, metrics, and noise model}
The swept axis is the \gls{rx} \gls{naf} separation $\Delta\ell^{(\mathrm{r})}$ of the two scatterers, with the uniform-weighting Rayleigh limit at $1/N_x^{(\mathrm{r})} \approx 0.09$. The \SI{45}{dB} Chebyshev baseline taper broadens this, so the swept range $[0.08, 0.28]$ reaches slightly more than twice the effective resolution. Detection follows the \gls{cfar} pipeline with iterative peak removal of~\cite{mandelli2022sampling} at $P^{(\mathrm{FA})} = 10^{-9}$. We report the missed-detection probability $P^{(\mathrm{MD})}$ and the \gls{naf}-\gls{rmse} of the localized peaks as in~\cite{felixOptimalAzimuthSampling2025a}, with the $F_1$ score including false alarms, over $10^4$ trials per $\Delta\ell^{(\mathrm{r})}$ point. \gls{awgn} at the \gls{rx} elements enters through the beamforming in~\eqref{eq:BeamformedSignalBistaticRx}, giving an image-plane SNR of $\approx \SI{17.6}{dB}$ on the oversampled methods, where
the references begin to lose detection, the noise-limited edge of the operating regime.

\paragraph{Noiseless regime}
All three methods saturate at parity in the noiseless regime (Fig.~\ref{fig:3d_eval_3way_merged}, top row), with every $P^{(\mathrm{MD})}$ curve below the $10^{-3}$ resolution of $2\times10^4$ trials from $\Delta\ell^{(\mathrm{r})} \approx 0.19$ on, except the sparsest isotropic reference from $0.21$, so the differentiator is acquisition cost, not detection quality. BASIIS reaches similar quality at $|\mathcal{B}| = 3441$ direction pairs, a $3.1\times$ and $5.3\times$ reduction relative to the densest isotropic and anisotropic oversampled points, with a \gls{naf}-\gls{rmse} below the isotropic references on the plateau, widening as they get sparser. This noiseless parity masks a difference the noisy regime exposes: BASIIS and the anisotropic reference share an elevation cell density the isotropic sampling lacks.

\paragraph{Operating-boundary regime (image SNR $\approx \SI{17.6}{dB}$)}
At the upper edge of the intended operating regime (Fig.~\ref{fig:3d_eval_3way_merged}, bottom row), BASIIS equals or outperforms the isotropic references on every metric, by a margin that widens as the interpolation factor decreases. The sparsest isotropic sampling at IF$=1.25$ degrades to $F_1 \approx 0.80$ at roughly three times the \gls{naf}-\gls{rmse} of BASIIS. The anisotropic reference acquires BASIIS's elevation cell density directly rather than reconstructing it from the coarray, so the two perform equally in $F_1$ while BASIIS concedes only a small $P^{(\mathrm{MD})}$ and \gls{naf}-\gls{rmse} margin that grows with the interpolation factor, where dense sampling proves marginally more robust at this noise level.

\paragraph{Reconstruction fidelity}
Fig.~\ref{fig:3d_recon} shows the noiseless \gls{rx}-pair \gls{naf} reconstruction, with BASIIS reproducing the dense reference response from the minimal acquisition. At the operating tolerance $\kappa = 10^{-3}$ the reconstruction is numerically indistinguishable from its $\kappa \to 0$ limit for this array pair at any baseline length, consistent with the bound of Appendix~\ref{ap:agcd}. The figure also includes the periodic-kernel variant. Resolving the finer grid yields the sharper noiseless image, without the direct kernel's sidelobe skirt, but fitting more detail than the $K$ samples support amplifies noise. The Monte Carlo evaluation therefore uses the direct kernel.

\paragraph{Acquisition cost}
The $3.1$ to $5.3\times$ fewer \gls{tx}--\gls{rx} direction pairs translate into radio-resource savings, as each pair costs one beamformer dwell: at one \acrshort{5G} NR slot per pair (\SI{0.5}{ms}), BASIIS acquires in \SI{1.7}{s} against \SI{5.3}{s} and \SI{9.2}{s} for the isotropic and anisotropic references, while its reconstruction, a few \glspl{dft} of order $K_{\max}$ per azimuth pair, costs orders of magnitude less.

\pgfplotsset{scaled y ticks=false}
\pgfplotsset{reconaxis/.style={
    scale only axis,
    width=0.23\columnwidth, height=0.25\columnwidth,   %
    axis on top, enlargelimits=false,
    xmin=-0.5, xmax=0.5, ymin=-0.5, ymax=0.5,
    xtick={-0.5,0,0.5}, ytick={-0.5,0,0.5},
    xticklabels={$\text{-}0.5$,$0$,$0.5$}, yticklabels={$\text{-}0.5$,$0$,$0.5$},
    xlabel={$\ell^{(\mathrm{r})}$}, ylabel={$\eta^{(\mathrm{r})}$},
    xticklabel style={/utils/exec={%
        \ifnum\ticknum=0\pgfkeysalso{xshift=4pt}\fi
        \ifnum\ticknum=2\pgfkeysalso{xshift=-4pt}\fi}},
    yticklabel style={/utils/exec={%
        \ifnum\ticknum=0\pgfkeysalso{yshift=4pt}\fi
        \ifnum\ticknum=2\pgfkeysalso{yshift=-4pt}\fi}},
    ylabel style={yshift=-1.3em}, xlabel style={yshift=0.2em},
    label style={font=\footnotesize}, tick label style={font=\footnotesize},
    colormap name=jet, point meta min=-60, point meta max=0}}
\newcommand{\recongfx}[1]{\addplot[forget plot] graphics[xmin=-0.5,xmax=0.5,ymin=-0.5,ymax=0.5,includegraphics={trim=1 0 0 1,clip}]{#1};}
\newcommand{\reconmarks}{\addplot[only marks, mark=x, color=white, mark options={line width=0.8pt}, mark size=2pt] coordinates {(-0.22,0.18) (0.20,-0.16)};}
\begin{figure}
  \centering
  \setcounter{subfigure}{0}%
  \begin{tikzpicture}
    \begin{groupplot}[reconaxis,
        group style={group size=3 by 1, horizontal sep=3mm,
                     y descriptions at=edge left}]   %
      \nextgroupplot
        \recongfx{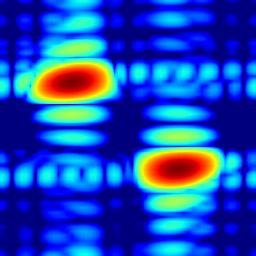}\reconmarks
      \nextgroupplot
        \recongfx{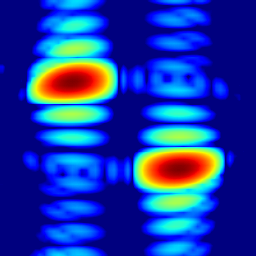}\reconmarks
      \nextgroupplot[colorbar,
          colorbar style={width=1.5mm, xshift=-2mm,
            ylabel={Norm.\ power \\ in dB},
            ylabel style={align=center, font=\scriptsize, yshift=3.0em},
            yticklabel style={font=\footnotesize},
            ytick={-60,-30,0}, yticklabels={$\text{-}60$,,$0$}}]
        \recongfx{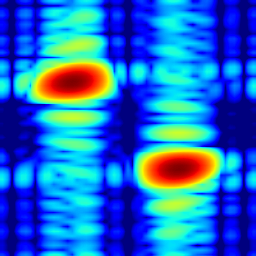}\reconmarks
    \end{groupplot}
    \coordinate (capy) at (current bounding box.south);
    \node[anchor=north, font=\footnotesize] at (group c1r1.south |- capy)
      {\refstepcounter{subfigure}(\thesubfigure)~Periodic\label{fig:3d_recon_periodic}};
    \node[anchor=north, font=\footnotesize] at (group c2r1.south |- capy)
      {\refstepcounter{subfigure}(\thesubfigure)~Reference\label{fig:3d_recon_reference}};
    \node[anchor=north, font=\footnotesize] at (group c3r1.south |- capy)
      {\refstepcounter{subfigure}(\thesubfigure)~Direct\label{fig:3d_recon_direct}};
  \end{tikzpicture}
  \vspace*{-1.5mm}
  \caption{Noiseless \gls{rx}-pair \gls{naf} reconstruction of two point scatterers ($\times$), max-projected over the \gls{tx} azimuth $\ell^{(\mathrm{t})}$: the oversampled reference (b) between BASIIS with the periodic (a) and the direct (c) Dirichlet kernel from the same minimal \gls{agcd} coarray acquisition.}
  \label{fig:3d_recon}
\end{figure}

\section{Conclusion}

We presented a complete angular sampling framework for bistatic \gls{isac}, extending the \gls{dft}-based optimal-sampling methodology of~\cite{mandelli2022sampling, felixOptimalAzimuthSampling2025a} from azimuth-only operation to the full \gls{4D} angular domain of the \gls{tx} and \gls{rx} arrays. The bistatic scan-on-scan constraint couples the two elevation angles, captured through the ortho-baseline coarray, whose aperture sets the joint elevation resolution per azimuth pair. From this we derived the Nyquist--Shannon-minimal set of \gls{tx}--\gls{rx} direction pairs together with a Dirichlet kernel reconstruction whose residual error is bounded by the \gls{agcd} tolerance. This delivers the minimal angular acquisition for any site separation, suited to practical beamforming operations in \gls{isac} deployments.

Monte Carlo simulations confirm that this minimal acquisition closely reproduces the dense oversampled reconstruction and essentially equalizes the detection accuracy of the oversampled references across the intended operating regime, requiring $3.1$ to $5.3\times$ fewer \gls{tx}--\gls{rx} direction pairs. Against the elevation-matched anisotropic reference it concedes only a small missed-detection and localization margin at the noise level shown, at similar detection $F_1$.
Future work includes rotated arrays ($k^\star$ then also depends on the orientation), multiple and extended targets, and measured bistatic data.

\appendices
\section{Approximate-GCD Coarray and Error Bound}
\label{ap:agcd}

The ortho-baseline coarray $\mathcal{C}$ of~\ref{sc:proposal} is uniform only where its \gls{naf}-unit generating spacings $g^{(\mathrm{t})} = k^{\star} d_z^{(\mathrm{t})}/\lambda$ and $g^{(\mathrm{r})} = d_z^{(\mathrm{r})}/\lambda$ share a common divisor\if\arxiv1, with exact grid spacing $d_{\mathcal{C}} = \gcd(g^{(\mathrm{t})}, g^{(\mathrm{r})})$\fi. For irrational $k^{\star}$ they are incommensurate and the exact \gls{gcd} is ill-conditioned.

We instead compute an \gls{agcd} at a finite tolerance $\kappa$ in \gls{naf} units: among the candidate spacings $g^{(\mathrm{r})}/q_i$ given by the continued-fraction convergents $p_i/q_i$ of $g^{(\mathrm{t})}/g^{(\mathrm{r})}$, each an exact divisor of $g^{(\mathrm{r})}$, we return as $\tilde{d}_{\mathcal{C}}$ the coarsest whose residual on $g^{(\mathrm{t})}$ stays within $\kappa$. Two safeguards share $k^{\star}_{\max} = 1/(2\kappa)$, which keeps the coarray spacing from falling below the tolerance itself: the translation factor is capped at $k^{\star}_{\max}$ where the scan geometry degenerates, keeping the aperture $L$ and the elevation order $K$ finite, while $\tilde{d}_{\mathcal{C}}$ is floored at $g^{(\mathrm{r})}/k^{\star}_{\max}$, bounding the ratio $g^{(\mathrm{r})}/\tilde{d}_{\mathcal{C}}$ and with it the period-matched kernel order.

The coarray spacing $\tilde{d}_{\mathcal{C}}$ is a near-divisor of both spacings, $g^{(\mathrm{t})} = p\,\tilde{d}_{\mathcal{C}} + \varepsilon^{(\mathrm{t})}$ and $g^{(\mathrm{r})} = q\,\tilde{d}_{\mathcal{C}} + \varepsilon^{(\mathrm{r})}$ with $p,q$ the nearest integers and $\varepsilon^{(\mathrm{t})}, \varepsilon^{(\mathrm{r})}$ the residual offsets. As the coarray depends only on the ortho-baseline element coordinates, we re-index $n,m$ over the $N_z^{(\hn\cdot\hn)}$ ortho-baseline coordinates $\mathcal{Z}^{(\hn\cdot\hn)}$ of~\eqref{eq:orthocoarray}, with $z_n^{(\mathrm{t})} = n\,d_z^{(\mathrm{t})}$ and likewise for the \gls{rx}. The coarray element $c_{n,m}/\lambda = n\,g^{(\mathrm{t})} + m\,g^{(\mathrm{r})}$ then lies at
\begin{equation}
    c_{n,m}/\lambda = (n p + m q)\,\tilde{d}_{\mathcal{C}} + \rho_{n,m},
    \label{eq:coarraymismatch}
\end{equation}
an integer multiple of $\tilde{d}_{\mathcal{C}}$ plus the residual mismatch $\rho_{n,m} = n\,\varepsilon^{(\mathrm{t})} + m\,\varepsilon^{(\mathrm{r})}$, whose worst case over the coarray indices obeys $\varepsilon_{\max} \leq (N_z^{(\mathrm{t})}-1)\,|\varepsilon^{(\mathrm{t})}| + (N_z^{(\mathrm{r})}-1)\,|\varepsilon^{(\mathrm{r})}|$, where $\varepsilon^{(\mathrm{r})} = 0$ and, above the floor, $|\varepsilon^{(\mathrm{t})}| \leq \kappa$ by construction, so $\varepsilon_{\max} \leq (N_z^{(\mathrm{t})}-1)\,\kappa$.

Snapping element $(n,m)$ onto its grid node perturbs its phasor $e^{-j2\pi (c_{n,m}/\lambda)\sin\varphi^{(\mathrm{r})}}$ by $\Delta\chi_{n,m} = 2\pi\,\rho_{n,m}\sin\varphi^{(\mathrm{r})}$, at most $2\pi\,\varepsilon_{\max}$ since $|\sin\varphi^{(\mathrm{r})}| \leq 1$. With $\hat{a}_{\kappa}$ the response reconstructed on the snapped coarray and $\hat{a}$ the true response of a unit-amplitude scatterer, both normalized to their peak gain\if\arxiv1, which leaves the array tapers unconstrained\fi, the non-negative tapers give $\sum_{n,m}|w_n^{(\mathrm{t})} w_m^{(\mathrm{r})}| = 1$, so with $|e^{j\Delta\chi}-1| \leq |\Delta\chi|$ the peak-relative error obeys
\begin{equation}
    \bigl|\hat{a}_{\kappa} - \hat{a}\bigr|
    \leq \sum_{n,m} |w_n^{(\mathrm{t})} w_m^{(\mathrm{r})}|\,|\Delta\chi_{n,m}|
    \leq 2\pi\, \varepsilon_{\max}.
    \label{eq:errboundraw}
\end{equation}
A finer tolerance $\kappa$ yields a finer coarray grid and smaller residuals, lowering the bound, which vanishes as $\kappa \to 0$. The reconstruction is lossless in that limit and near-lossless at the finite tolerance in practice, quantifying the claim of~\ref{sc:proposal}.

\section*{Acknowledgment}
This work was developed within the SENSation project, partly funded by the Federal Ministry of Research, Technology and Space under grant 16KIS2523K and 16KIS2532.

\balance

\bibliographystyle{IEEEtran}
\bibliography{references.bib}

\end{document}